# Spin-wave Confinement and Coupling in Organic-based Magnetic Nanostructures

Michael Chilcote[1], Megan Harberts[1], Bodo Fuhrmann[2], Katrin Lehmann[3], Yu Lu[4], Andrew Franson[1], Howard Yu[1], Na Zhu[5], Hong Tang[5], Georg Schmidt[2,3], and Ezekiel Johnston-Halperin[1,a]

[1]Department of Physics, The Ohio State University, Columbus, OH 43210-1117, USA

[2]IZM, Martin-Luther-Universität Halle-Wittenberg, Halle, 06120, Germany

[3]Institute für Physik, Martin-Luther-Universität Halle-Wittenberg, Halle, 01620, Germany

[4]Department of Chemistry, The Ohio State University, Columbus, OH 43210-1173, USA

[5]Department of Electrical Engineering, Yale University, New Haven, Connecticut 06511, USA

[a]Author to whom correspondence should be addressed. Electronic address: johnston-halperin.1@osu.edu

**Abstract:** Vanadium tetracyanoethylene (V[TCNE]$_x$) is an organic-based ferrimagnet that exhibits robust magnetic ordering ($T_C$ of over 600 K), high quality-factor (high-$Q$) microwave resonance ($Q$ up to 3,500), and compatibility with a wide variety of substrates and encapsulation technologies. Here, we substantially expand the potential scope and impact of this emerging material by demonstrating the ability to produce engineered nanostructures with tailored magnetic anisotropy that serve as a platform for the exploration of cavity magnonics, revealing strongly coupled quantum confined standing wave modes that can be tuned into and out of resonance with an applied magnetic field. Specifically, time-domain micromagnetic simulations of these nanostructures faithfully reproduce the experimentally measured spectra, including the quasi-uniform mode and higher-order spin-wave (magnon) modes. Finally, when the two dominant magnon modes present in the spectra are brought into resonance by varying the orientation of the in-plane magnetic field, we observe anti-crossing behavior indicating strong coherent coupling between these two magnon modes at room temperature. These results position V[TCNE]$_x$ as a leading candidate for the development of coherent magnonics, with potential applications ranging from microwave electronics to quantum information.

The recent success of organic-based thin films in the areas of optoelectronics and electronics promises a new materials basis for these applications that is mechanically flexible, facile to synthesize, and low cost when compared to traditional inorganic materials.[1–3] This success should in principle extend to magnetic and spintronic functionality, and to some extent this promise has been realized in the observation of spin-dependent phenomenology including organic magnetoresistance[4–7] (OMAR), organic magneto-electroluminescence[8,9] (OMEL), spin-pumping and spin transport,[10] and related phenomena.[11–13] However, this phenomenology is constrained by the fact that spins in these materials exhibit only diamagnetic, or at best paramagnetic, ordering and therefore miss the rich phenomenology found in extended magnetic order (such as ferro- and ferrimagnetism). In particular, applications in the emerging field of coherent magnonics rely implicitly on the ability to excite and exploit long lived spin wave excitations in a magnetic material. That requirement has led to the nearly universal reliance on yttrium iron garnet (YIG), which requires epitaxial synthesis on lattice matched substrates at temperatures above 800 °C to achieve high materials quality[14–16] and has reigned for half a century as the unchallenged leader in low loss magnetic resonance despite extensive efforts to identify alternative materials. Surprisingly, organic-based ferrimagnets of the form M[Acceptor]$_x$ (M = transition metal; x ≈ 2) provide one of the most promising routes to realizing this goal, with the room-temperature, low-loss ferrimagnet vanadium tetracyanoethylene (V[TCNE]$_x$) emerging as a compelling alternative to YIG. Manifestations of the potential of this material system can be found in the demonstration of control of magnetic properties via ligand-tuning[17–19] and metal-substitution,[20–25] optimized synthesis[26] ($T_C$ > 600 K), extremely sharp (typically 1 Oe at 9.86 GHz) ferromagnetic resonance (FMR) features,[27,28] the demonstration of FMR-driven spin-pumping,[29] and encapsulation strategies that stabilize the magnetic properties for weeks to months under ambient conditions.[30]

Here we build on this recent progress to demonstrate the ability to control the morphology of V[TCNE]$_x$ magnetic structures by creating arrays of templated nanowires, yielding control of magnetic anisotropy and resulting spin-wave mode coupling and quantum confinement with no substantial increase in damping. This control is achieved through growth on SiO$_2$ substrates patterned with nanoscale grooves using ultraviolet interference lithography. After growth, these nanowire structures exhibit a high quality-factor (high-$Q$) quasi-uniform ferromagnetic resonance (FMR) mode with uniaxial crystal-field driven magnetic anisotropy of 23.527 Oe ± 0.083 Oe, oriented perpendicular to the nanowire axis. We perform time-domain micromagnetic simulations of these nanostructures to provide additional insight into the mode structure present in the experimentally measured spectra. With the results, we identify the two dominant magnon modes present in the spectra: one mode results from the resonant excitation of the magnetic material in the nanowire itself, while the other stems from the resonant excitation of the nanostructured magnetic material found within the trenches that lie between the wires. When these two magnon modes are brought into resonance by varying the orientation of an in-plane magnetic field, we observe anti-crossing behavior consistent with strong, coherent coupling between the two modes. This study positions V[TCNE]$_x$ as a leading candidate for the development of coherent magnonics, with functionality that directly challenges the best *inorganic* thin films demonstrated to date.[14–16]

V[TCNE]$_x$ samples are synthesized using a previously reported chemical vapor deposition (CVD) growth process.[26] Figure 1a shows a schematic view of a custom-built CVD reactor that is housed within an argon glovebox. During the deposition, argon gas carries the two precursors, TCNE and V(CO)$_6$, into the reaction zone (shaded green in Fig. 1a) where V[TCNE]$_x$ is deposited onto one or more substrates. The system employs three independently temperature-controlled regions for the TCNE, V(CO)$_6$, and reaction zone with typical setpoints of 70 °C, 10 °C, and 50 °C, respectively. For this experiment, patterned $SiO_2$ substrates are prepared using ultraviolet interference lithography and reactive ion etching to produce an alternating pattern of trenches and ridges. A variety of patterned substrates were prepared, with pitches varying from 200 nm to 350 nm and ridge widths varying from 63 nm to 180 nm. All growth runs consist of deposition onto one or more patterned substrates as well as a control sample consisting of either a flat $SiO_2$ or sapphire wafer die to account for any growth-to-growth variation in $V[TCNE]_x$ thin film properties.

The result of a typical growth on the $SiO_2$ templates described above can be seen in Fig. 1b, where cross-sectional scanning electron microscopy (SEM) shows the silicon wafer (dark grey), the patterned $SiO_2$ layer (light grey), and the CVD deposited $V[TCNE]_x$ layer (dark grey) from the bottom to the top of the image, respectively. This microscopy reveals that the film growth around the high aspect ratio features are governed by well-known growth dynamics controlled in part by the differential arrival angles of gas-phase precursors around the features, which results in thicker coverage on exterior angles than interior angles.[31] Over the course of the deposition, $V[TCNE]_x$ forms into wire-like structures sitting atop the $SiO_2$ ridges and leaves closed-off voids within the trenches (Fig. 1b). Studies of substrates oriented with trenches parallel and perpendicular to the gas flow direction in the CVD furnace reveal no significant changes in morphology or magnetic characteristics between the two orientations.

After growth, samples are mounted in the appropriate orientation and sealed in electron paramagnetic resonance (EPR) grade quartz tubes without exposure to air. When not being measured, the samples are stored in a -35 °C freezer within an argon glovebox and are found to be stable for weeks. The results of room-temperature DC magnetometry measurements as a function of applied magnetic field for a nanowire sample are shown in Fig. 1c. In contrast to the case for uniform thin films of V[TCNE]$_x$, the magnetization response to an in-plane field depends on whether that field is applied parallel or perpendicular to the trench axis. This behavior is consistent with the formation of an easy axis aligned perpendicular to the trenches (open triangles), showing saturation at a lower field than when the applied field is perpendicular to the wires (hard axis, filled squares).

While these results are suggestive, a more complete study of the magnetic anisotropy in these V[TCNE]$_x$ nanostructures can be found via FMR characterization of the anisotropy fields. Room-temperature measurements are made using a Bruker electron paramagnetic resonance spectrometer, configured with an X-band bridge with 200 µW of applied microwave power and a modulation field of 0.05 Oe. In standard operation, the microwave frequency is tuned between 9 and 10 GHz for optimal microwave cavity performance before the measurement, and then the frequency is fixed while the DC field is swept during the measurement. For consistency, the same template material as that used for the SQUID measurement shown in Fig. 1c was used for the FMR studies shown here. Figure 2a shows the FMR spectra of a V[TCNE]$_x$ nanowire array for the magnetic field applied in-plane ($\theta = 90°$; see inset to Fig. 2f) as the sample is rotated for values of $\phi$ ranging from -90° to 270°. Figure 2b shows the integrated microwave absorption, as opposed to the synchronously detected derivative spectra shown in Fig. 2a. These spectra show two sets of features, each 90° phase shifted from each other and covering a different field range. At angles where the external field is applied perpendicular to the trench (i.e. $\phi = $ -90°, 90°, 270°), both high and low field features contain two peaks, suggesting the presence of higher order confined spin wave excitations supported by the nanostructured V[TCNE]$_x$. This multimodal behavior persists through a subset of the full angular range shown in Figs. 2a and 2b. However, at high symmetry angles where the external field is applied parallel to the trench (i.e. $\phi = $ 0°, 180°), there is an isolated, single peaked feature as well as a peak with much lower amplitude at lower field. Most strikingly, at angles where these two sets of features would in principle cross (approximately at $\phi$ = -45°, 45°, 135°, and 225°) a gap appears in the spectra. This is most clearly apparent in the integrated spectra shown in Fig. 2b and in Fig. 2c, which shows the extracted center field values for the higher intensity set of peaks. Both show an avoided crossing with a 14 Oe gap.

The peak with the highest intensity is ascribed to the quasi-uniform FMR mode of the nanowires as it involves the largest volume of magnetic material in the sample (this assumption will be validated by the analysis below). The angular variation shown in Fig. 2c suggests a uniaxial anisotropy with an easy axis perpendicular to the nanowire/trench axis. Interestingly, this outcome is contrary to what one might expect from a simple magnetostatics argument using the shape anisotropy of a long thin rod (which would predict an anisotropy field of approximately 50 Oe with an easy magnetization axis parallel to the wire axis), and therefore further detailed analysis is necessary to understand the origin of the magnetic anisotropy present in these nanostructures. In order to expand on the observations above, additional measurements are performed for rotations from in-plane to out of plane both for orientations parallel ($\phi$ = 0°) and perpendicular ($\phi$ = 90°) to the nanowire axis with the results shown in Figs. 2d and 2e. The center fields for the quasi-uniform mode are extracted and all three data sets are simultaneously fit to the same set of equations (solid and dashed lines in Figs. 2c and 2f).

Due to the complex geometry present in these samples many of the simplifying assumptions typically employed in fitting thin magnetic films do not apply, as a result the data are fit to dispersions obtained from the complete magnetic free energy of the nanowire array. The magnetic free energy, including contributions from the Zeeman energy, an effective anisotropy energy, and an uniaxial anisotropy energy, is expressed as[32,33]

$$F = -\boldsymbol{M} \cdot \boldsymbol{H} + \tfrac{1}{2}M[(\boldsymbol{m} \cdot \boldsymbol{H}_{\text{eff}})^2/H_{\text{eff}} - (\boldsymbol{m} \cdot \boldsymbol{H}_\parallel)^2/H_\parallel], \quad (1)$$

where $\boldsymbol{M}$ is the magnetization vector, $\boldsymbol{m}$ is the magnetization unit vector given by $\boldsymbol{M}/M$, $\boldsymbol{H}$ is the applied bias field vector, $\boldsymbol{H}_{\text{eff}}$ is an effective field resulting primarily from shape anisotropy (see discussion following Eq. 4), and $\boldsymbol{H}_\parallel$ is a uniaxial anisotropy field, which points in the direction of the easy magnetization axis. The first term in Eq. 1 corresponds to the Zeeman energy while the second term includes contributions from the demagnetization energy and uniaxial anisotropy energy. In general, the effective field $\boldsymbol{H}_{\text{eff}}$ is not in the same direction as uniaxial anisotropy field $\boldsymbol{H}_\parallel$. For the present case, the data presented in Fig. 2 indicate that $\boldsymbol{H}_\parallel$ is in-plane, perpendicular to the nanowire axis (i.e. along $\theta = 90°$, $\phi = 90°$). The shape anisotropy of the nanowire array, given its high packing fraction, is similar to that of a thin film, and so $\boldsymbol{H}_{\text{eff}}$ is oriented normal to the surface of the sample (i.e. $\theta = 0°$) as is typically the case for thin films. Note that $\boldsymbol{H}_{\text{eff}}$ includes contributions from any perpendicular uniaxial crystal-field anisotropy present in the sample since both this anisotropy and the demagnetization energy have an identical angular dependence, and therefore are indistinguishable in FMR studies. For the analysis presented here these additional contributions to the anisotropy energy are not explicitly considered.

The resonance frequency, $\omega$, can then be determined using the formalism provided by Smit, Beljers, and Suhl,[34,35]

$$\omega = \frac{\gamma}{M \sin\theta}(F_{\theta\theta}F_{\phi\phi} - F_{\theta\phi}^2)^{1/2} \quad (2)$$

where $\gamma$ is the gyromagnetic ratio and $F_{ij}$ is the second derivative of the free energy $F$ with respect to the angles $i$ and $j$. The FMR resonance fields (Fig. 2) are more than an order of magnitude larger than the typical saturation field for V[TCNE]$_x$, and therefore we assume that the magnetization is effectively parallel to the applied magnetic field (i.e. $\phi \approx \phi_H$ and $\theta \approx \theta_H$ where $\theta$, $\phi$ and $\theta_H$, $\phi_H$ are the polar and azimuthal angles of the magnetization $\boldsymbol{M}$ and the applied bias field $\boldsymbol{H}$, respectively). With this framework, we may now separately obtain the dispersion relation for each sample orientation in Fig. 2 using Eqs. 1 and 2:

$$\frac{\omega}{\gamma} = \sqrt{(H - H_\parallel \cos 2\phi_H)(H + H_{\text{eff}} + H_\parallel \sin^2 \phi_H)}, \quad (\theta_H = 90°) \quad (3a)$$

$$\frac{\omega}{\gamma} = \sqrt{(H - H_\parallel - H_{\text{eff}} \cos^2 \theta_H)(H - H_{\text{eff}} \cos 2\theta_H)}, \quad (\phi_H = 0°) \quad (3b)$$

$$\frac{\omega}{\gamma} = \sqrt{\frac{1}{2}(H - (H_\parallel + H_{\text{eff}})\cos 2\theta_H)(2H + H_\parallel - H_{\text{eff}} - (H_\parallel + H_{\text{eff}})\cos 2\theta_H)}, \quad (\phi_H = 90°) \quad (3c)$$

The first equation (Eq. 3a) is for an in-plane rotation ($\theta_H = 90°$) as the applied field is rotated through $\phi_H$. The second and third equation are for in-plane to out of plane rotations where the field is applied either along the nanowire axis (Eq. 3b; $\phi_H = 0°$) or perpendicular to the nanowire axis (Eq. 3b; $\phi_H = 90°$) as the applied field is rotated through $\theta_H$. Using an alternate form of the Smit-Beljers-Suhl formula,[36] this set of equations may be written as a single equation. This relation applies for arbitrary rotations of the applied field through both $\theta_H$ and $\phi_H$ and reduces to Eqs. 3a–c given the appropriate constraints. For rotations along the symmetry axes of the geometry presented here (i.e. for $\theta_H$ rotations along $\phi_H = 0°$, 90°, 180°, 270° and for $\phi_H$ rotations along $\theta_H = 90°$; or more formally, when $\cos \theta_H \sin 2\phi_H = 0$), this relation reduces to:

$$\frac{\omega}{\gamma} = \sqrt{\frac{1}{2}(H - (H_{\text{eff}} + H_\parallel \sin^2 \phi_H)\cos 2\theta_H)(2(H - H_\parallel \cos 2\phi_H) - (H_{\text{eff}} + H_\parallel \sin^2 \phi_H) - (H_{\text{eff}} + H_\parallel \sin^2 \phi_H)\cos 2\theta_H)}. \quad (3d)$$

As noted above $\phi \approx \phi_H$ and $\theta \approx \theta_H$, and so from here forward we will drop the subscript, using simply $\phi$ and $\theta$.

When all three data sets are simultaneously fit (solid and dashed lines in Figs. 2c and 2f; the open data points shown in Fig. 2c, which are affected by the presence of the avoided crossing, are excluded from the fitting process), this set of equations allows for the self-consistent extraction of $H_{\text{eff}}$ and $H_\parallel$, yielding values of 91.188 Oe ± 0.510 Oe and 23.527 Oe ± 0.083 Oe, respectively. This value of $H_{\text{eff}}$ is consistent with previous reported values of $4\pi M_S$ for uniform thin films; for example, $4\pi M_S$ of 95 Oe is reported in Reference 27. In considering the origin of the uniaxial anisotropy field, $H_\parallel$, it is important to note that the shape anisotropy from magnetostatic effects in a long thin rod would create an easy axis parallel to the nanowire axis, rather than the anisotropy perpendicular to the nanowire axis observed in Fig. 2. As a result, $H_\parallel$ must arise from a crystal field anisotropy wherein the local exchange vector acquires some anisotropy due to either lattice symmetry or strain.

The difference in the coefficients of thermal expansion for organic and inorganic materials often vary by an order of magnitude or more, and have been reported to affect electronic properties of the organic materials.[37,38] As a result, it is likely that an anisotropic strain field is created in the nanowire structures due to the continuous contact with the $SiO_2$ substrate along the nanowire axis and the ability for the nanowires to relax along the radial direction due to the presence of the grooves. This phenomenology, along with successful fitting using Eq. 3, suggests that the higher intensity set of peaks in Fig. 2 do result from the quasi-uniform FMR mode of the nanowire and that the easy axis is indeed, surprisingly, perpendicular to the patterning axis. We also note that the magnetic material within the trenches is not subjected to the anisotropic strain field induced by the ridges in the nanowires, and therefore becomes a leading candidate for the second set of resonance peaks present in Fig. 2.

Further insight can be gained by comparing these results to thin films grown on unpatterned substrates, wherein no in-plane anisotropy is observed (see supplementary material). Previously, this lack of in-plane anisotropy has been interpreted to mean no anisotropy fields exist. However, this data suggests an alternative explanation. In films deposited as a uniform thin film, strain due to differential thermal expansion would be applied uniformly in the plane of the film, yielding significantly simpler, modified forms of Eq. 3:

$$\frac{\omega}{\gamma} = \sqrt{H\left(H + (4\pi M_S - H_A)\right)}, \quad (\theta = 90°) \tag{4a}$$

in which the $\phi$ dependence has dropped out. And for the full in-plane to out of plane rotation,

$$\frac{\omega}{\gamma} = \sqrt{(H - (4\pi M_S - H_A)\cos^2\theta)(H - (4\pi M_S - H_A)\cos 2\theta)}. \quad (4b)$$

Here, we have included the contribution from the demagnetization fields of a thin film as $4\pi M_S$ and have conformed to the sign convention typical of $H_A$, which arises from uniformly applied strain in the plane of the film. Note that while this reproduces the lack of in-plane anisotropy observed for thin films, it implies an additional anisotropy field in the out of plane direction. Coincidentally, this anisotropy field has the same symmetry, but not necessarily the same sign, as the shape anisotropy for a thin film. This in turn implies that prior measurements of the anisotropy of thin films are in fact measuring $4\pi M_S - H_A = 4\pi M_{eff}$ rather than the bare $4\pi M_S$ as previously assumed.[27,28] In the literature, $H_A$ often presents itself as $H_\perp$ because it is responsible for inducing perpendicular anisotropy. In the present study, the width of the ridges on which the nanowires are templated is not zero, and therefore there may be some residual in-plane strain perpendicular to the nanowire axis generating a residual anisotropy field, $H_A$. However, as with previous studies of uniform thin films, there is no straightforward way to disentangle this residual anisotropy from $4\pi M_S$, leading us to use the more general $H_{eff} \equiv 4\pi M_{eff}$ in defining Eq. 3.

While this analysis resolves several long-standing mysteries in the nature of magnetic ordering and anisotropy in V[TCNE]$_x$, it only describes the primary peak visible in Fig. 2 and does not describe either the additional resonances or the anti-crossing behavior noted above. In order to answer these questions, the effective field analysis described in Eqs. 1–3 is used to inform quantitative time-domain micromagnetic simulations using the open-source GPU-accelerated simulation software MuMax3.[39] The results of the simulations are shown in Fig 3, with the geometry determined by the real structure of the nanowires as extracted from the corresponding SEM images (Figs. 3e–j). Figure 3a shows the in-plane experimental FMR data previously shown in Fig. 2b, while Fig. 3b shows a plot of the simulated FMR data over the same field range as the integrated spectra. Figures 3c–d show experimental spectra compared directly to the corresponding simulated spectra for geometries with the field applied perpendicular ($\phi = 90°$) and parallel ($\phi = 0°$) to the nanowire array. The fit values of $H_{eff}$ and $H_\parallel$ extracted from the dataset (Fig. 2) and literature values for the Gilbert damping constant and exchange constant[28,29] are used as the materials parameters input into the simulation. The simulations are run using an out of plane continuous-wave microwave excitation, as was the case in the real experiment.[40,41] Consistent with the phenomenology proposed above, the magnetic material in the nanowires themselves are simulated with a uniaxial magnetic anisotropy. However, the material in the trenches does not include this anisotropy and instead, only includes contributions from the demagnetizing fields. Additionally, the simulations were found to most faithfully reproduce the experimental data when the top and bottom surfaces are (perfectly) pinned.

Figures 3e–j show resonant microwave excitation mode maps created by overlaying the change in the z-component of the reduced magnetization ($\Delta m_z$) onto the SEM micrograph used in the simulation. The colored bezel around each mode map corresponds to the color of the dashed line in Fig 3c or d to which the mode map corresponds. Each panel shows the structure of the resonant excitations supported by the $V[TCNE]_x$ array at the indicated position on the corresponding spectra.

The simulation results reveal that the two sets of peaks in the data are the result of a quasi-uniform mode supported in the nanowire and a second resonantly excited mode supported by the magnetic material within the trenches of the $SiO_2$ template, consistent with the initial assumptions above. Furthermore, as the angle of the field is rotated from perpendicular to the wires (e.g. $\phi = -90°$) to parallel to the wires (e.g. $\phi = 0°$), the excitation mode structure hybridizes as the two dominant peaks come together. At $\phi = 50°$ in the simulated data (see Fig. S1 in the supporting material), a node forms between what was formerly a pure trough mode excitation and the quasi-uniform FMR mode of the nanowire. These two intimately linked but spatially distinct regions exist in substantially different magnetic environments; they represent two high-Q magnon cavities, each with its own magnetic anisotropy, connected by a continuous low-damping magnetic material. As a result, the two cavities can be tuned into and out of resonance with each other using an applied magnetic field, and when their individual resonant conditions approach the point where they would coincide, the modes hybridize, and the result is an avoided crossing in the FMR data.

The gap between the mode branches in this regime is 14 Oe, corresponding to an energy ($\mu_B B$) of 0.081 μeV, while the half-width of the gap in terms of a frequency ($\gamma$) corresponds to a spacing of 20 MHz. This gap is approximately seven times the peak-to-peak linewidth and 10% of the full field variation of the quasi-uniform mode, indicating that these two excitations are in the strong coupling regime.[42–45] While there is also a significant change in the intensity of these lines as they proceed through the crossing, it is difficult to disentangle effects due to the intrinsic strength of these resonances from the efficiency of their detection due to complicating factors such as the fact that this data is acquired by the physical rotation of the sample within the microwave cavity (which can perturb the cavity mode) and the varying spatial symmetries of the modes (which can affect their coupling efficiency to the microwave cavity and therefore their detection).

In conclusion, this work demonstrates the ability to engineer the magnetic anisotropy in thin films deposited on patterned substrates and to engineer both the dispersion and anisotropy of confined spin wave modes in templated V[TCNE]$_x$ nanowires. Nanowires with a diameter of approximately 300 nm are grown on the plateaus between grooves, exhibit the high-Q quasi-uniform FMR mode, and display anisotropy with a shift in resonant field of 23.527 Oe ± 0.083 Oe.

Finally, when the trough spin-wave mode and quasi-uniform mode are brought into resonance by varying the orientation of an in-plane magnetic field, we observe anti-crossing behavior and the opening of a gap of 14 Oe, indicating strong coherent coupling between these two excitations at room temperature. These results position V[TCNE]$_x$ as a leading candidate for the development of coherent magnonics, with potential applications ranging from microwave electronics[28] to quantum information.[46–49]

**EXPERIMENTAL SECTION**

The samples in this study consist of organic-based magnetic nanostructures of vanadium tetracyanoethylene (V[TCNE]$_x$) that assemble along the ridges of a grooved substrate. To fabricate the grooved SiO$_2$ substrates, 35 nm of Cr was sputtered onto SiO$_2$(1 µm)/Si(100) wafers and subsequently coated with photoresist in preparation for laser interference lithography. The samples were exposed using a custom, home-built laser interference lithography setup equipped with a 266 nm laser and then etched using an Oxford Plasmalab 100 system ICP 180 to produce a regular alternating pattern of trenches and ridges with pitch ranging across different samples from 200 nm to 350 nm and ridge width varying from 63 nm to 180 nm. The resist was then stripped using a standard O$_2$ plasma clean and the Cr mask was removed using the reactive ion etching system previously noted.

The V[TCNE]$_x$ layer was grown in a custom chemical vapor deposition (CVD) setup via the reaction of vanadium hexacarbonyl (V(CO)$_6$) with tetracyanoethylene (TCNE) in argon carrier gas. The precursors were prepared according to standard techniques in the literature and the nanowire samples were synthesized using the same CVD growth process that has previously been optimized for thin films.

All samples were mounted so as to prevent unwanted rotation and sealed in evacuated electron spin resonance (ESR) grade quartz tubes immediately after growth and without exposure to air. When not being measured, the sealed samples were stored in a -35 °C freezer within an argon glovebox. When necessary, samples were manipulated and remounted within an argon glovebox before being resealed in quartz tubes for additional measurements.

Superconducting quantum interference device (SQUID) magnetometry measurements were collected using a Quantum Design Magnetic Property Measurement System using the Reciprocating Sample Option (RSO) and with an applied field of 100 Oe. Cavity ferromagnetic resonance (FMR) measurements were made using an X-band Bruker ESR spectrometer at room temperature with an applied microwave power of 200 µW. The microwave frequency was tuned between 9 GHz and 10 GHz for optimal microwave cavity performance before starting the measurement.

All of the FMR data was collected from a single high-quality 3 mm × 3 mm sample (patterned with a nominal pitch of 300 nm and trench width of 170 nm), from which the four edges were cleaved to avoid spurious effects introduced by V[TCNE]$_x$ growth on the substrate edges; the cleaved edges were used for the cross-sectional scanning electron microscopy seen in Figs. 3 and S1. The SQUID data shown in Fig. 1c was collected on a separate sample grown on a substrate cleaved from the same template as that used for Figs. 2 and 3 (again, with a nominal pitch of 300 nm and trench width of 170 nm). The micrograph in Fig. 1b shows a nanostructured V[TCNE]$_x$ growth on a template with a nominal pitch of 350 nm and a nominal trench width of 220 nm.

Micromagnetic simulations were performed using the open-source GPU-accelerated micromagnetic simulation software MuMax3. The simulations were performed using the real geometry of the nanostructures as extracted from SEM micrographs of the sample, with periodic boundary conditions applied in order to simulate an array. The simulations were run using a continuous wave approach, with an out of plane continuous-wave microwave excitation, as was the case in the real experiment. Except for the Gilbert damping constant (for which an artificially high value was used to reduce simulation time; $\alpha = 8.0 \times 10^{-4}$) and the exchange constant (for which literature values were used), all materials parameters used in the simulation were obtained from fitting the experimental FMR data. The magnetic material on top of each $SiO_2$ ridge was modeled to include a contribution from the uniaxial anisotropy obtained by fitting the FMR data. The magnetic material within the trenches was modeled without the inclusion of any additional anisotropy beyond standard contributions from the demagnetizing fields. The top and bottom surfaces were perfectly pinned during all simulations. The mode maps in Fig. 3 show the change in the z-component of the reduced magnetization ($\Delta m_z$) overlaid on SEM micrographs.

## SUPPLEMENTARY MATERIAL

Supplementary material, available from AIP Publishing or the corresponding author, contains additional micromagnetic simulation mode maps of the quasi-uniform mode, the full equation for the angular dependence obtained from the alternate form of the Smit-Beljers-Suhl equation, thin-film control data, additional data on a sample with a thicker magnetic layer, and data on the dependence of the magnetic anisotropy as a function of both the thickness of the deposited $V[TCNE]_x$ layer and the pitch of the underlying $SiO_2$ templated substrate.

## ACKNOWLEDGMENTS

This work was supported by NSF Grant No. DMR-1507775 and DFG Grant No. SFB762. The authors acknowledge the NanoSystems Laboratory at Ohio State University, the high-performance cluster computing facilities provided by The Ohio State University Arts and Sciences Technology

Services, and technical assistance from and discussion with Dr. Rohan Adur, Dr. Shane White, Dr. William Ruane, and Dr. Michael Flatté.17

**Figure 1.** (a) Plan view of the chemical vapor deposition schematic. (b) Cross-sectional scanning electron micrograph of V[TCNE]$_x$ nanowires grown on a patterned SiO$_2$ substrate. The inset shows an alternate view of the same sample demonstrating that the nanowires extend the length of the sample (scale bar: 1 µm). (c) Magnetization as a function of applied field at room temperature for V[TCNE]$_x$ grown on a patterned substrate. The inset shows the coordinate system used across all experiments, with the nanowires aligned parallel to the x-axis. The magnetization is normalized by the saturation magnetization and plotted for $H_{app}$ perpendicular ($H_{app}^\perp$; $\theta = 90°$, $\phi = 90°$) to the substrate pattern (△) and parallel ($H_{app}^\parallel$; $\theta = 90°$, $\phi = 0°$) to the substrate pattern (■).

**Figure 2.** (a) Shows FMR spectra for the magnetic field applied in-plane as the sample is rotated for values of $\phi$ ranging from -90° to 270° (shown from bottom to top and as labeled on the right side of the data) for a V[TCNE]$_x$ nanowire array. (b) Shows the integrated microwave absorption (numerical integration of the data as shown in (a)), as opposed to the synchronously detected derivative spectra. (c) Shows the extracted center fields of the in-plane angular series shown in (a) and (b). The solid line is a fit to the data (see text). Panels (d) and (e) show the FMR spectra for rotations from in-plane to out of plane both for orientations parallel ($\phi = 0°$) and perpendicular ($\phi = 90°$) to the nanowire axis, respectively. (f) Shows the extracted center fields from the angular series shown in (d) and (e) with fits shown as solid and dashed lines. The inset shows the coordinate system with respect to the sample geometry.

**Figure 3.** (a) Shows the integrated in-plane experimental FMR data previously shown in Fig 2b. (b) Shows a plot of the simulated FMR data over the same field range for values of $\phi$ ranging from -90° to 270° (shown from bottom to top and as labeled on the right side of the data) for a V[TCNE]$_x$ nanowire array. The spectrum simulated for 0° is shown in red, and the spectrum simulated for 90°

is shown in blue. (c) Shows the experimental spectrum collected at 90° in the top panel and the simulated spectrum for 90° in the bottom panel in red. The four dashed lines in the bottom panel correspond to the mode maps shown in (e), (f), (g), and (h). (d) Shows the experimental spectrum collected at 0° in the top panel and the simulated spectrum for 0° in the bottom panel in blue. The two dashed lines in the bottom panel correspond to the mode maps shown in (i) and (j). (e)--(j) Show mode maps overlaid on the SEM micrograph used in the simulation (see text; the change in the z-component of the reduced magnetization, $\Delta m_z$, is shown). The colored bezel around each mode map corresponds to the color of the dashed line in (c) or (d) to which the mode map corresponds. Scale bar: 500 nm.

**Figure 1**

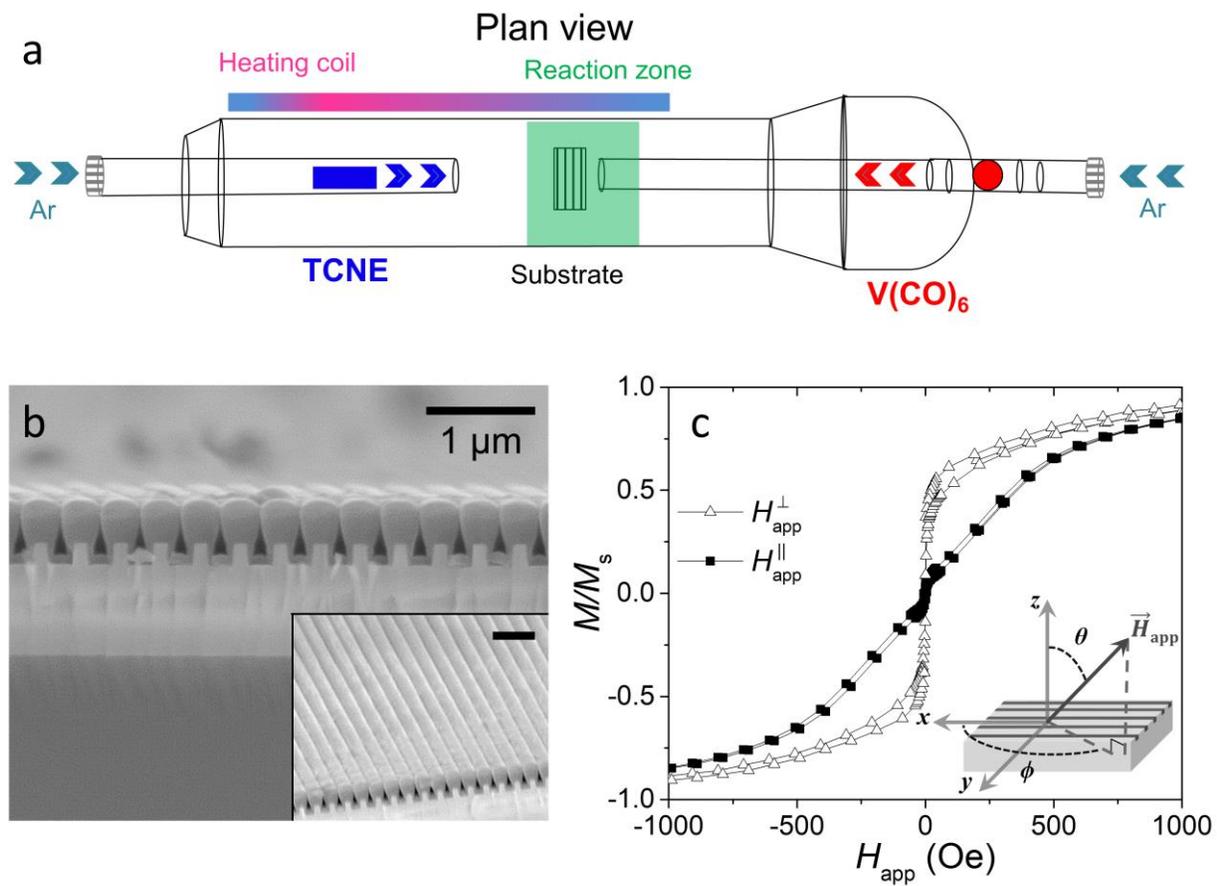

**Figure 2**

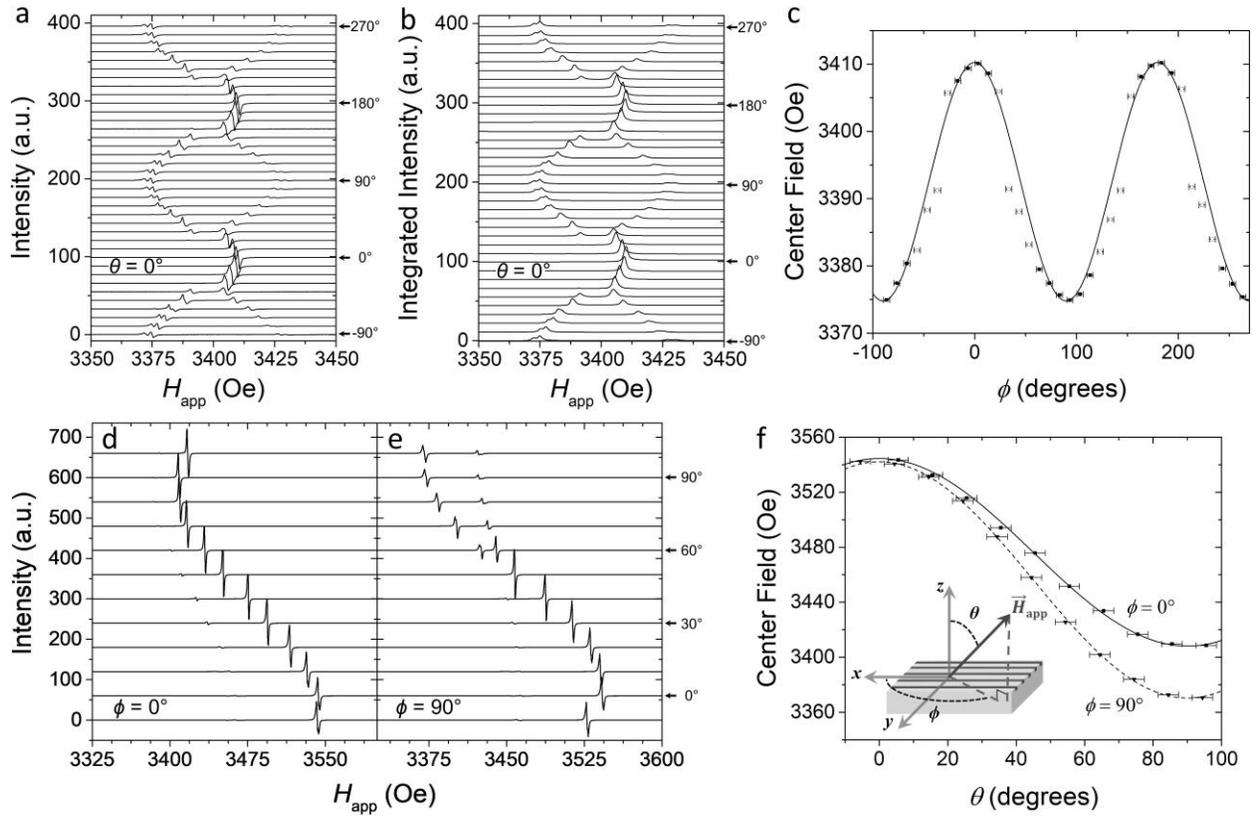

**Figure 3**

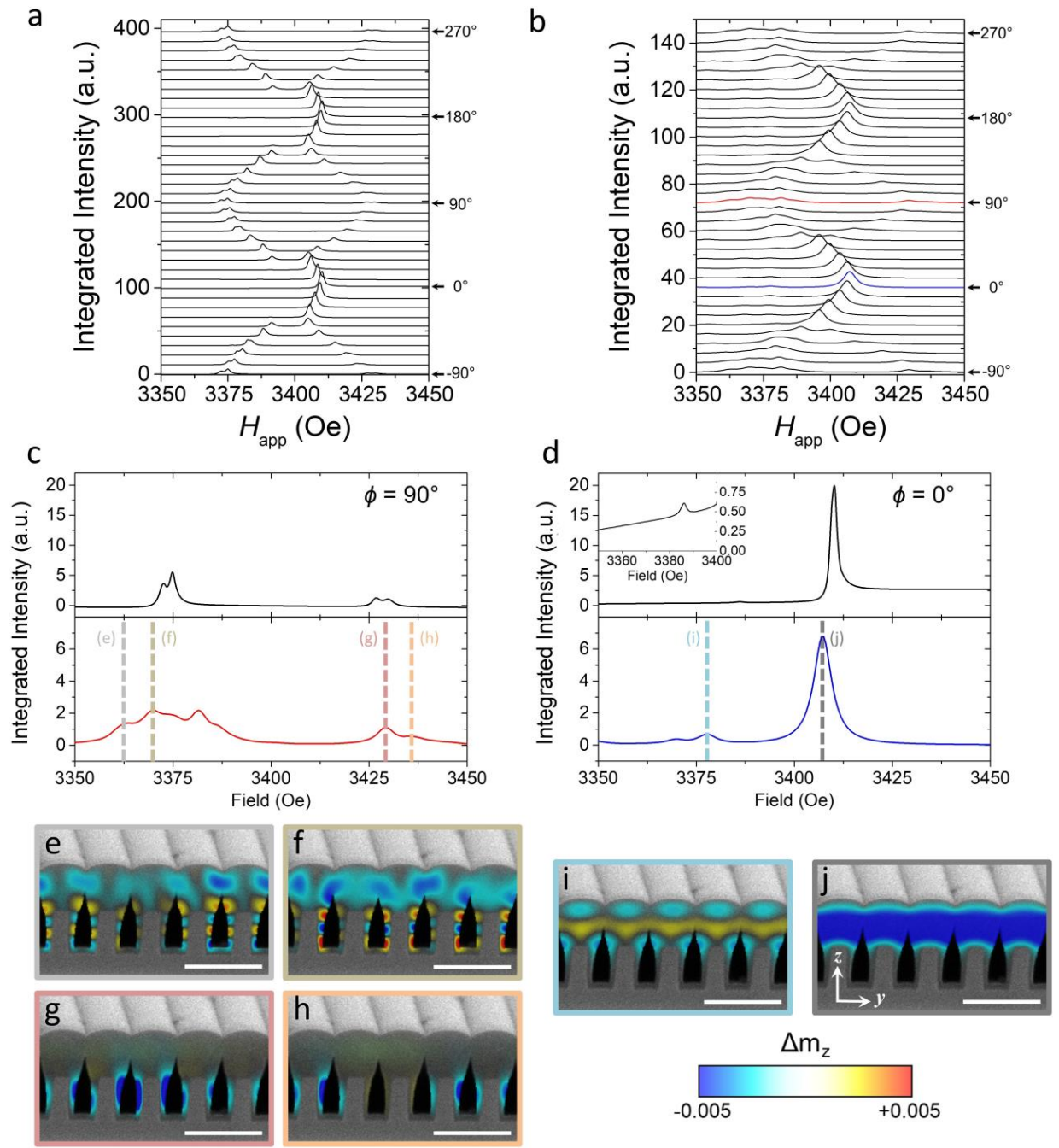